\documentclass[a4paper,11pt]{article}
\usepackage{pos}

\title{Matching lattice QC$+$ED to Nature}

\author*{Nazario Tantalo}

\affiliation{University of Rome Tor Vergata and INFN sezione di Roma Tor Vergata,\\
  Via della Ricerca Scientifica 1, Rome, Italy}

\emailAdd{nazario.tantalo@roma2.infn.it}

\abstract{The first step of any QFT calculation, aiming at phenomenological predictions, is the matching of the theory to Nature. The matching procedure fixes the parameters of the theory in terms of an equal number of external inputs that, if the theory is expected to reproduce observations, must be experimentally measured physical quantities. 
At the sub-percent level of accuracy QED radiative corrections become important and the theory expected to describe the hadronic Universe is QCD+QED. Phenomenological predictions deriving from lattice QCD calculations do depend, at this level of precision, upon the choice of the external inputs used to match/define the approximate theory.
In this written version of my talk I concentrate on the theoretical aspects of the matching procedure of lattice QCD$+$QED and of the definition of QCD, strongly advocating a community agreement on the matching scheme to be used in future lattice QCD calculations. 
}

\FullConference{%
  The 39th International Symposium on Lattice Field Theory (Lattice2022),\\
  8-13 August, 2022 \\
  Bonn, Germany 
}

\begin{document}
\maketitle

\section{Introduction}
Lattice QCD calculations entered the precision era. This statement is certainly true but we need to elaborate a bit on it to better appreciate its meaning and, more importantly, in order to understand its consequences. 

By precision we mean sub-percent accuracy and by lattice QCD calculations we mean a restricted set of hadronic observables, the so-called gold-plated quantities. There are many interesting hadronic observables, such as those associated with processes involving more than a single hadron in the initial and/or the external state, on which much more theoretical and numerical work is needed in order to reach even the ten-percent level of accuracy. At the same time, a restricted set of hadronic observables has been computed, by more than one lattice collaboration, with an overall accuracy, statistical plus systematics, at the permille-level of accuracy (see Ref.~\cite{FlavourLatticeAveragingGroupFLAG:2021npn} for a recent review). Gold-plated observables include stable hadron masses and also phenomenologically very relevant quantities such as the renormalized strong coupling constant, renormalized quark masses, leptonic and semileptonic decay rates of pseudoscalar mesons, and so on.      

The price that has to be payed in order to live in the precision era, i.e.\ to further improve the accuracy on gold-plated quantities, is the inclusion of QED radiative corrections and strong isospin breaking (SIB) effects in non-perturbative lattice calculations. In fact, the theory expected to describe the hadronic universe at the sub-percent level of accuracy is QCD$+$QED. Consequently, QCD has to be considered an approximation at this level of precision, although an excellent one. 

The necessity of performing lattice simulations of QCD$+$QED has been, by now, fully recognized by the lattice community. Starting from the pioneering simulations of Ref.~\cite{Duncan:1996xy}, a huge amount of work has been done in order to cope with the subtle numerical and theoretical issues associated with the inclusion of electromagnetic interactions in finite-volume lattice simulations. A critical and exhaustive discussion of all these works goes far beyond the scope of this talk (see Refs.~\cite{Blum:2007cy,Hayakawa:2008an,Blum:2010ym,deDivitiis:2011eh,deDivitiis:2013xla,Budapest-Marseille-Wuppertal:2013rtp,Davoudi:2014qua,Borsanyi:2014jba,Lucini:2015hfa,Horsley:2015vla,Horsley:2015eaa,Endres:2015gda,Carrasco:2015xwa,Lee:2015rua,Tantalo:2016vxk,Lubicz:2016xro,Blum:2017cer,Giusti:2017dmp,Patella:2017fgk,Boyle:2017gzv,Bussone:2017xkb,Feng:2018qpx,Davoudi:2018qpl,MILC:2018ddw,Bijnens:2019ejw,CSSM:2019jmq,Giusti:2019xct,Hatton:2020qhk,DiCarlo:2021apt,Feng:2021zek,Hatton:2020lnm,Aoki:2012st,FermilabLattice:2021hzx} for a largely incomplete list of references). Therefore, in the following, I concentrate on the main subjects of this talk, i.e. the theoretical aspects of the matching of lattice QCD$+$QED to Nature and of the scheme ambiguities arising in the definition of QCD as an approximate theory.

By working under the assumption that QCD$+$QED is the fundamental theory, its matching to Nature is unambiguous and the choice of the experimental observables to be used as inputs of the matching procedure is a matter of numerical convenience. On the contrary, by considering QCD as an approximate theory, different choices of the external inputs used to tune the parameters of the theory correspond to different schemes for the definition of what we call QCD. At the sub-percent level of precision QCD results and QED radiative corrections  obtained in different schemes cannot be compared. Since (slightly) different schemes have been adopted in the past by the different lattice collaborations~\cite{FlavourLatticeAveragingGroupFLAG:2021npn}, the issue is becoming increasingly relevant and has attracted a lot of attention (see for example Refs.~\cite{Gasser:2003hk,deDivitiis:2013xla,Budapest-Marseille-Wuppertal:2013rtp,Bussone:2018ybs,DiCarlo:2019thl,Borsanyi:2020mff,Sachrajda:2021enz,Boyle:2022lsi} and also the contribution of A.~Portelli at this conference).

\section{Setting the stage}
\label{sec:thematching}
\begin{figure}[t!]
\begin{center}
\includegraphics[width=0.8\textwidth]{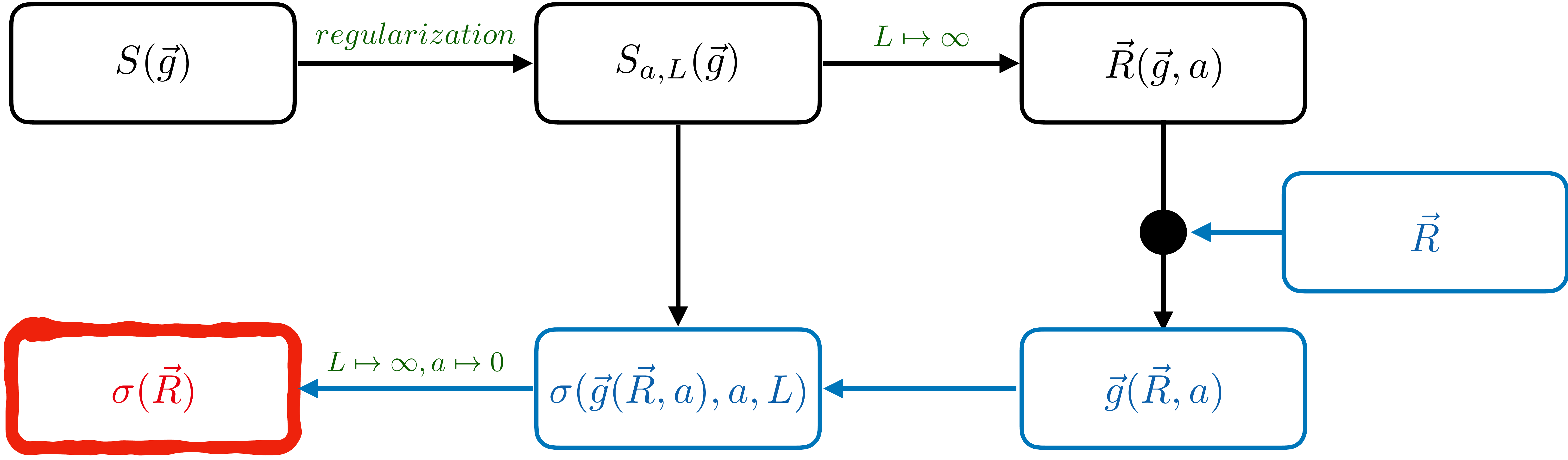}
\end{center}
\caption{\label{fig:algorithm} Illustration of the matching algorithm in the case of a generic theory.}
\end{figure}
The first step of any quantum field theory (QFT) calculation, aiming at phenomenological predictions, is the matching of the theory to Nature. A renormalizable theory depending upon $N$ bare parameters can describe $\mathbb{R}^N$ Universes. The matching ``algorithm'', also known as \emph{renormalization}, selects the one in which we live. The algorithm is illustrated in Figure~\ref{fig:algorithm} in the case of a \emph{generic} theory and we are now going to describe it:
\begin{itemize}

\item we start (top-left rectangle) with the classical action $S(\vec g)$ of our theory, depending upon $N$ bare parameters collected into the vector $\vec g$; 

\item the theory needs to be regularized and we do it by going on the lattice. The lattice action $S_{a,L}(\vec g)$ (top-center rectangle) depends upon the lattice spacing $a$ and the finite volume $L$ that regularize the theory, respectively, in the ultraviolet (UV) and in the infrared (IR);

\item we can now (assuming that we have access to a supercomputer) start our calculations. By making educated guesses, we chose some values for our $N$ bare parameters, fix the number of lattice points, run the simulations and compute a collection of observables. Let's assume that our final goal is the calculation of a particular quantity that we call $\sigma$ (bottom-left red rectangle). For each choice of the vector $\vec g$ we thus need to compute $\sigma$ but also $N$ additional observables that we are not going to predict but that we will use to tune the bare parameters. We collect these observables into the vector $\vec R$ and, at this stage, our theoretical results depend upon the choices that we have made, i.e. we have $\sigma(\vec g,a,L)$ and $\vec R(\vec g,a,L)$;

\item here we work under the assumption that our theory is the fundamental one, i.e.\ it is expected to describe our Universe, and therefore impose the matching condition (black blob),
\begin{flalign}
\lim_{L\mapsto \infty} \vec R\left(\vec g,a,L\right) = \vec R\;,
\label{eq:matchinga}
\end{flalign}
by using the experimental values for $\vec R$ on the r.h.s.\ (blue rectangle on the right of the blob). If instead we want to select another Universe, we just need to choose arbitrary values for $\vec R$;

\item we now solve the matching conditions w.r.t.\ the bare parameters at fixed UV cutoff, thus obtaining $\vec g(\vec R,a)$. We could live with IR-dependent bare couplings but this is a bit odd and also unpractical and therefore we have taken the infinite-volume limit in Eq.~(\ref{eq:matchinga});

\item we now evaluate our theoretical prediction at the matching point, i.e.\ $\sigma(\vec g(\vec R,a),a,L)$, iterate the procedure above by \emph{choosing} increasingly smaller values of the lattice spacing and, finally, obtain our prediction for $\sigma$,
\begin{flalign}
\lim_{L\mapsto \infty,a\mapsto 0} \sigma\left(\vec g(\vec R,a),a,L\right) = \sigma(\vec R)\;.
\label{eq:continuuma}
\end{flalign}
\end{itemize}

As already stressed, this is just renormalization theory and, therefore, there is nothing new in the discussion of this section. In fact almost nothing is new in the whole talk. Nevertheless, because of the peculiarities of QCD$+$QED and of the way we usually do things on the lattice, there are some delicate points on which it might be helpful to elaborate a bit more. We will do this in the following subsections.

\subsection{Renormalized couplings, theory scales, lines of constant physics and all that}
\label{sec:afmatching}
Up to now we have been discussing renormalization theory without mentioning renormalized couplings. The external inputs have been used, at fixed UV cutoff, to tune directly the bare parameters. This is what we usually do on the lattice. 
On the other hand, renormalized couplings play a prominent r\^ole in any textbook discussion of renormalization theory and, in fact, can be very useful also on the lattice, at least from the practical/numerical point of view. 

There is no conceptual difference between the so-called theory scales and the renormalized couplings. These are quantities that cannot be directly measured in experiments but that can be defined and  computed theoretically. In fact, if we can compute a theoretical quantity on the lattice with high accuracy we can then profitably use it to define the so-called \emph{lines of constant physics}. 

The important point to be noticed is that the very same Universe can be selected by using different, but consistent, external inputs. Let's assume that we use $\vec R$ to match our theory and then, together with $\sigma$, we predict $N$ additional observables
\begin{flalign}
\vec R \quad \longrightarrow \quad \vec g(\vec R,a) \quad \longrightarrow \quad 
\left\{
\sigma(\vec R)\;,
\
\vec S=\vec S(\vec R)
\right\} \;.
\end{flalign}
If instead of $\vec R$ we now use $\vec S$ to match the theory we have
\begin{flalign}
\vec S \quad \longrightarrow \quad \vec g(\vec S,a) \quad \longrightarrow \quad 
\left\{
\sigma(\vec S)\;,
\
\vec R(\vec S)
\right\} \;.
\end{flalign}
In any healthy and consistent theory we must have $\vec R(\vec S)=\vec R$ and, therefore, the same line of constant physics, $\sigma(\vec S)=\sigma(\vec R)$. Of course, if the theory is expected to describe \emph{our} Universe then we must have $\vec R(\vec S^\mathrm{exp})=\vec R^\mathrm{exp}$. If this doesn't happen we have discovered new physics. On the other hand, once QCD is considered an approximation of the real Universe, lattice QCD results cannot be expected to reproduce the experimental observations and to be independent from the matching observables. This dependence is at the origin of the scheme ambiguities arising in the definition of QCD and we will discuss them in section~\ref{sec:approximation}. 

Let's now assume that we can compute $N$ quantities that have a well defined continuum and infinite-volume limit but that cannot be directly measured in experiments. For example, we might want to compute renormalized quark masses $\hat m_f(\mu)$ in a given scheme and at a given scale $\mu$, the renormalized fine structure constant $\hat \alpha_\mathrm{em}(\mu)$, the gradient-flow scales $t_0$~\cite{Luscher:2010iy} and $w_0$~\cite{Borsanyi:2012zs}, unphysical meson masses ($M^\Gamma_{\psi\psi}$) extracted from the fermion-connected Wick contractions of two-point Euclidean correlators of $\bar \psi \Gamma \psi$ interpolating operators, and so on. Let's collect these $N$ quantities in a vector that we call $\vec G$. The idea here is to start exploring unphysical Universes by prescribing arbitrary values for $\vec G$, to take the continuum limit and in the end to match our theory to Nature by dealing with UV-finite quantities. This is actually the way renormalization is explained in QFT textbooks. To do this we first tune the bare parameters at fixed $\vec G$,
\begin{flalign}
\lim_{L\mapsto \infty} \vec G\left(\vec g,a,L\right) = \vec G
\quad\longrightarrow\quad \vec g(\vec G,a)\;.
\label{eq:matchingg}
\end{flalign}
This defines the lines of constant physics corresponding to the selected values of $\vec G$. We than take the continuum limit on these lines,
\begin{flalign}
\lim_{L\mapsto \infty,a\mapsto 0} \vec R\left(\vec g(\vec G,a),a,L\right) = \vec R(\vec G)\;,
\qquad
\lim_{L\mapsto \infty,a\mapsto 0} \sigma\left(\vec g(\vec G,a),a,L\right) = \sigma(\vec G)\;.
\label{eq:continuumg}
\end{flalign}
In this way we are trading the dependence of the measurable observables $\vec R$ and $\sigma$ upon the bare parameters (at fixed UV cutoff) with the dependence upon $\vec G$ (in the continuum) and, as a matter of fact, the quantities $\vec G$ act as renormalized couplings. At this point we can impose the matching conditions in the continuum by solving
\begin{flalign}
\vec R(\vec G) = \vec R^\mathrm{exp}
\quad\longrightarrow\quad \vec G(\vec R^\mathrm{exp})\;,
\label{eq:continuumgmatching}
\end{flalign}
and reach our final goal, i.e. to predict $\sigma(\vec R^\mathrm{exp})\equiv \sigma(\vec G(\vec R^\mathrm{exp}))$. 

The following important observation, being rather obvious, is usually left implicit. The tuning of the renormalized parameters with experimental inputs has to be done, once in history at least! It is this step that matches the theory to our Universe. It is this step that removes all the ambiguities, associated with the choice of the observables $\vec G$, that seem to be present in the renormalization procedure. If someone, once upon a time, solved Eq.~(\ref{eq:continuumgmatching}) for us, we can then select the line of constant physics corresponding to our Universe by using its solution $\vec G(\vec R^\mathrm{exp})$ in the matching of our lattice simulations. This might be particularly convenient from the numerical viewpoint, depending on how precisely we are able to compute the chosen theoretical observables $\vec G$.

\subsection{Lattice QCD$+$QED and asymptotic freedom}
We now consider \emph{our} theory. 
 
The bare parameters of QCD$+$QED are the strong coupling constant $g_s$, the (squared) electric charge $e^2$ and the $n_f$ quark masses, where $n_f$ is the number of dynamical flavours. At the current level of precision, as far as low-energy observables are concerned, the effects associated with propagating bottom and top quarks can safely be neglected. Therefore, by simulating on the lattice QCD$+$QED with dynamical up, down, strange and charm quarks, any observable will depend upon six bare parameters as well as on the lattice spacing and on the lattice volume. 
 
In order to avoid lengthy expressions, such as $[aM_p](g_s,e^2,am_u,am_d,am_s,am_c,a,L)$ where $aM_p$ is the proton mass and $am_f$ are the bare quark masses expressed in lattice units, we could use the vector of the bare couplings $\vec g=(g_s,e^2,am_u,am_d,am_s,am_c)$ introduced in the previous sections. In the discussion that follows, however, it is more convenient to introduce   
\begin{flalign}
\vec x=\left(e^2,am_u,am_d,am_s,am_c,a\right)
\label{eq:xdef}
\end{flalign}
and use the compact notation $[aM_p](g_s,\vec x,L)$. 

In section~\ref{sec:thematching} we emphasized that in order to implement the matching algorithm we need to \emph{chose} increasingly smaller values for the lattice spacing. This might sound quite strange to a novice lattice practitioner (Simplicio), used to think that the lattice spacing has to be \emph{determined}. In fact a moment of thought reveals that nothing prevent us from choosing the value of the UV cutoff. Although by working in lattice units we have in our hands dimensionless quantities, say $[aM_p](g_s,\vec x,L)$, once the value of the lattice spacing has been chosen, say $1/a=4$~GeV, these quantities can easily be expressed in physical units, $M_p(g_s,\vec x,L)=4\times [aM_p](g_s,\vec x,L)$~GeV, and used in the matching conditions, with $M_p^\mathrm{exp}$, to determine $\vec g(\vec R,a)$. This is certainly possible but (for numerical convenience?) we usually prefer to follow a different route.

\begin{figure}[t!]
\begin{center}
\includegraphics[width=0.8\textwidth]{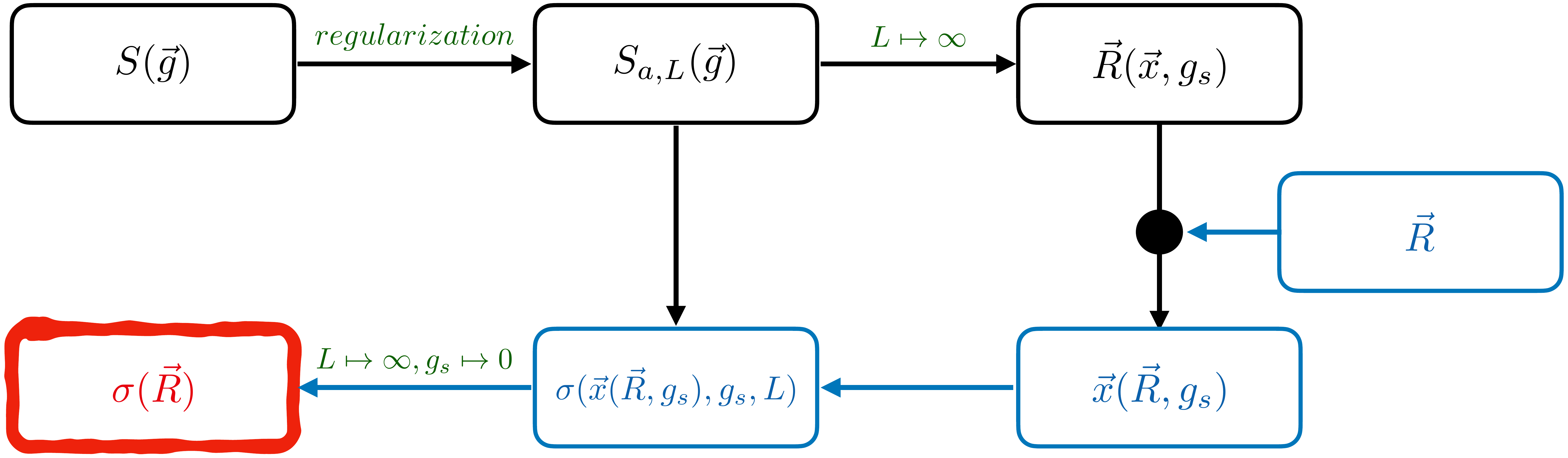}
\end{center}
\caption{\label{fig:algorithm2} Illustration of the matching algorithm in the case of an asymptotically free theory in which the bare coupling $g_s$ vanishes in the $a\mapsto 0$ limit. The other bare couplings and the lattice spacing are collected into the vector $\vec x$, see Eq.~(\ref{eq:xdef}).}
\end{figure}
By relying on asymptotic freedom, instead of choosing the value of the lattice spacing we prefer to choose the value of the bare strong coupling $g_s$ . We then use the matching conditions to fix the remaining bare parameters \emph{and} the lattice spacing,
\begin{flalign}
\lim_{L\mapsto \infty} \vec R\left(g_s,\vec x,L\right) = \vec R\;,
\quad \longmapsto \quad \vec x(\vec R,g_s)\;.
\label{eq:matchinggs}
\end{flalign}
In this modified matching procedure, that is possible in the first place only because strong interactions are asymptotically free (see Figure~\ref{fig:algorithm2}), the lattice spacing is actually determined. Increasingly smaller values of the lattice spacing are obtained by relying on the fact that 
\begin{flalign}
\lim_{g_s\mapsto 0} a(\vec R,g_s) = 0\;,
\end{flalign}
and theoretical predictions are finally obtained by taking the continuum and infinite-volume limits according to
\begin{flalign}
\lim_{L\mapsto \infty,g_s\mapsto 0} \sigma\left(g_s,\vec x(\vec R,g_s),L\right) = \sigma(\vec R)\;.
\label{eq:continuumgs}
\end{flalign}
%

\section{How do we chose the matching observables?}
Under the assumption that the theory is the fundamental one, QCD$+$QED in our case, the choice of the observables to be used in the matching conditions is a matter of computational convenience. 

Let's go back to Eq~(\ref{eq:matchinggs}). Both the experimental inputs (r.h.s.) and the lattice calculations (l.h.s.) are unavoidably affected by errors. These errors propagate on the tuned bare parameters and on the lattice spacing, $\vec x(\vec R,g_s)$, and in the end on any theoretical prediction. The experimental precision is usually not a problem in this game. The theoretical precision might instead be an issue and, therefore, we are used to chose the matching observables among the hadronic quantities that can be computed on the lattice easily and precisely. 

At this point Simplicio immediately thinks to pseudoscalar meson masses, to the leptonic decay constants of charged pions ($f_\pi$) and kaons ($f_K$) and to the masses of the nucleons and of the $\Omega^-$ baryons. 
These are indeed the quantities that have been commonly adopted in the matching procedure of QCD to Nature under the assumption that QED and SIB corrections are negligible, see Ref.~\cite{FlavourLatticeAveragingGroupFLAG:2021npn}. In particular, $f_\pi$ and $f_K$ have usually been preferred to baryon masses because of the well known problem of the exponential degradation of the signal-to-noise ratio at large Euclidean times affecting baryon correlators.

The situation needs to be reconsidered in QCD$+$QED. By turning-on electromagnetic interactions massless particles appear into the spectrum, the photons, and observables that we are used to compute precisely and efficiently in QCD might become much more cumbersome in QCD$+$QED. 

This is certainly the case of decay rates. Propagating virtual photons generate $\log(L)$ infrared-divergent terms at intermediate stages of the calculation of $S$-matrix elements that, according to the well-known Bloch and Nordsieck mechanism~\cite{Bloch:1937pw}, cancel in the measurable observable by also considering processes with real photons in the final state. The infrared-safe measurable quantity associated with the process $\pi^+\mapsto \mu^+\nu_\mu(\gamma)$ is the decay rate $\Gamma\left[\pi^+\mapsto \mu^+\nu_\mu(\gamma),E_\gamma \right]$, inclusive on the number of real photons in the final state with total energy $E_\gamma$. The quantity that in QCD$+$QED corresponds to what we call $f_\pi$ in QCD (in the limit in which electromagnetic interactions are turned-off) is
\begin{flalign}
\mathcal{F}_\pi(E_\gamma)=
\sqrt{
\frac{\Gamma\left[\pi^+\mapsto \mu^+\nu_\mu(\gamma),E_\gamma \right]}{
\frac{G_F^2}{8\pi}\vert V_{ud}\vert^2 M_{\pi^+}^\mathrm{exp} (M_\mu^\mathrm{exp})^2\left[
1-\left(\frac{M_\mu^\mathrm{exp}}{M_{\pi^+}^\mathrm{exp}}\right)^2
\right]
}}\;,
\end{flalign}
where $G_F$ is the Fermi constant and $V_{ud}$ is the CKM matrix element. On the one hand, $\mathcal{F}_\pi(E_\gamma)$ and $\mathcal{F}_K(E_\gamma)$ can be~\cite{Carrasco:2015xwa,Lubicz:2016xro,Tantalo:2016vxk,DiCarlo:2021apt} and \emph{have been}~\cite{Giusti:2017dwk,DiCarlo:2019thl,Boyle:2022lsi} computed on the lattice. On the other hand, since the calculation is much more involved than in the QCD case and, more importantly, since prior knowledge of the relevant CKM matrix elements is required, I'm firmly convinced that using $\mathcal{F}_\pi(E_\gamma)$ and/or $\mathcal{F}_K(E_\gamma)$ as matching observables is a very bad idea!

Stable hadron masses are infrared safe quantities. Nevertheless, finite-volume effects on stable charged hadron masses are suppressed only as inverse powers of the volume in QCD$+$QED while they vanish exponentially fast in QCD. In the so-called QCD$+$QED$_\mathrm{L}$ and QCD$+$QED$_\mathrm{C}$ finite-volume formulations of the theory it is possible, by using the formulae of Ref.~\cite{Borsanyi:2014jba,Lucini:2015hfa}, to remove analytically the leading $O(e^2/L)$ and $O(e^2/L^2)$ finite-volume effects that are universal, i.e.\ depend only on the electric charge and on the mass of the hadron but not on its internal structure. The remaining non-universal structure-dependent finite-volume effects have to be removed by performing a numerical extrapolation (the $L\mapsto \infty$ limit appearing in Eq.~(\ref{eq:matchinggs})). 

There is another important point, again associated with the presence of photons into the spectrum, that must be taken into account concerning the calculation of stable hadron masses in lattice QCD$+$QED. In QCD stable hadrons are true eigenstates of the Hamiltonian, also in the infinite volume. Let's consider a charged pion. At vanishing spatial momentum, $M_{\pi^+}$ is the eigenvalue of the QCD Hamiltonian and, in this channel, the continuum part of the spectrum starts at $M_{\pi^+}+2M_{\pi^0}$. In QCD$+$QED, since this is a flavored channel, we still have a gap from the vacuum but, because of the presence of states such as $\pi^+ +\gamma$, the continuum part of the spectrum starts exactly at $M_{\pi^+}$. This fact, on the long run, will have important implications on the way we usually compute hadron masses on the lattice, i.e. the so-called effective-mass analysis. Indeed, while in lattice QCD the charged-pion effective-mass converges at large Euclidean times to $M_{\pi^+}$ with leading corrections proportional to $\exp(-2M_{\pi^0}t)$, in QCD$+$QED the corrections are $\exp(-E_\gamma^\mathrm{min}t)$ where $E_\gamma^\mathrm{min}\sim 1/L$ is different from zero only because of the finite-volume quantization of the spectrum. On the volumes that have been explored so far numerical evidences that stable hadron masses can be precisely computed in lattice QCD$+$QED have been provided in many works (see for example the current state-of-the-art calculation of the baryon spectrum~\cite{Borsanyi:2014jba} and Refs.~\cite{Hansen:2018zre,Bushnaq:2022aam} were, although at unphysical quark masses, a fully gauge-invariant approach has been used). On asymptotically large volumes, approaches based on spectral-reconstruction techniques~\cite{DelDebbio:2022uve,Hansen:2019idp} might turn helpful in this game. 

In light of the previous observations, two examples of possible choices for the matching observables are given by
\begin{flalign}
&
\vec R=\left(
\frac{M_{\pi^+}^2}{M_{K^0}^2},
\frac{M_{D^+}^2}{M_{K^0}^2},
\frac{M_{D_s}^2}{M_{K^0}^2},
M_{K^0},
\frac{M_{K^+}^2-M_{K^0}^2}{M_{K^0}^2},
\frac{M_{D^+}^2-M_{D^0}^2}{M_{K^0}^2}
\right) \;,
\nonumber \\
\nonumber \\
&
\vec R=\left(
\frac{M_{\pi^+}^2}{M_{\Omega^-}^2},
\frac{M_{K^0}^2}{M_{\Omega^-}^2},
\frac{M_{D_s}^2}{M_{\Omega^-}^2},
M_{\Omega^-},
\frac{M_{K^+}^2-M_{K^0}^2}{M_{\Omega^-}^2},
\frac{M_{D^+}^2-M_{D^0}^2}{M_{\Omega^-}^2}
\right)\;.
\label{eq:fullRs}
\end{flalign}
In both cases, as customary, we have considered dimensionless ratios and a single dimensional quantity that, in the lattice jargon, is the one that sets the scale. The mass differences $M_{K^+}^2-M_{K^0}^2$ and $M_{D^+}^2-M_{D^0}^2$, that vanish in the isospin symmetric limit, are the more sensitive quantities to the isospin breaking (IB) bare parameters $e^2$ and $am_u-am_d$. 
In the first case baryon masses are avoided at the price of inducing a strong dependence of light quark masses and of the lattice spacing upon charmed meson masses and, thus, potentially large cutoff effects. The second case is the one that, perhaps, can be considered the natural choice but, maybe, it could be better to consider a different baryon, such as the proton or the $\Xi$ (see e.g. Refs.~\cite{RBC:2014ntl,Borsanyi:2020mff,Segner:2022dou,RQCD:2022xux})). This discussion can be continued forever\dots and therefore we stop it here!

\section{QCD is an approximation}
\label{sec:approximation}
QCD and QCD$+$QED are two different theories, both in the IR and in the UV. 

In the previous section we have been mostly concerned with the different IR behaviors. 
In this section we are going to address the issue of the \emph{definition} of QCD as an approximate theory and we need to take in due consideration the fact that the two theories have different UV divergences. The issue becomes subtle because isospin symmetric QCD (isoQCD) is an excellent approximation of the hadronic Universe and, therefore, QED and SIB effects on physical quantities are small. This is not automatic though, it depends on how we define the bare parameters of (iso)QCD that, because of the different UV divergences, are infinitely different from the corresponding ones of QCD$+$QED.

In order to avoid lengthy expressions we need, again, to introduce a convenient notation. Let's introduce the vectors
\begin{flalign}
\vec y=\left(a,\frac{am_u+am_d}{2},am_s,am_c\right)\;,
\qquad
\vec z=\left(\frac{am_u-am_d}{2},e^2\right)\;.
\label{eq:splittingbare}
\end{flalign}
At fixed UV and IR cutoffs, any QCD$+$QED observable, say $\sigma$, depends upon the volume $L$, the bare strong coupling constant $g_s$, the lattice spacing $a$, the bare iso-symmetric light quark mass $am_{ud}=(am_u+am_d)/2$, the strange and charm bare masses as well as on the IB light quark mass difference $(am_u-am_d)/2$ and the bare electric charge squared $e^2$, i.e.\ $\sigma(g_s,\vec y,\vec z,L)$. The same observable in isoQCD is obtained by setting to zero the IB bare parameters collected into the vector $\vec z$, i.e.\ $\sigma(g_s,\vec y,\vec 0,L)$. 

Let's now assume that in order to match QCD$+$QED we choose, for the needed six experimental inputs, the second line of Eqs.~(\ref{eq:fullRs}) and let's split the corresponding vector $\vec R$ as follows
\begin{flalign}
&
\vec R=\left(\vec R_0, \vec R_1\right)\;,
\qquad
\vec R_0=\left(
\frac{M_{\pi^+}^2}{M_{\Omega^-}^2},
\frac{M_{K^0}^2}{M_{\Omega^-}^2},
\frac{M_{D_s}^2}{M_{\Omega^-}^2},
M_{\Omega^-}
\right)
\qquad
\vec R_1=\left(
\frac{M_{K^+}^2-M_{K^0}^2}{M_{\Omega^-}^2},
\frac{M_{D^+}^2-M_{D^0}^2}{M_{\Omega^-}^2}
\right)\;.
\label{eq:RR0R1}
\end{flalign}
At this point we have all the ingredients to discuss the definition of isoQCD and of the associated QED and SIB corrections.

Let's start by considering again QCD$+$QED but, this time, let's implement the matching algorithm of subsection~\ref{sec:afmatching} by splitting it into two separate steps, according to
\begin{flalign}
&
\lim_{L\mapsto \infty}\vec R_0\left(g_s,\vec y,\vec z,L\right) = \vec R_0\;,
\quad \longmapsto \quad \vec y(\vec R_0,g_s,\vec z)\;,
\label{eq:twosteps1}
\\
&
\lim_{L\mapsto \infty}\vec R_1\left(g_s,\vec y(\vec R_0,g_s,\vec z),\vec z,L\right) = \vec R_1\;,
\quad \longmapsto \quad 
\left\{
\begin{array}{l}
\vec z(\vec R,g_s)
\\
\\
\vec y\left(\vec R_0,g_s,\vec z(\vec R,g_s)\right)
\end{array}
\right.\;.
\label{eq:twosteps2}
\end{flalign}
In the first step, Eq.~(\ref{eq:twosteps1}), we solve the system of four equations at fixed values of $g_s$ and of the IB bare couplings $\vec z$, thus obtaining the lattice spacing and the three iso-symmetric quark masses $\vec y(\vec R_0,g_s,\vec z)$. In the second step, Eq.~(\ref{eq:twosteps2}), we complete the matching by using the solution of the first step and by solving the remaining two equations w.r.t.\ the IB bare couplings. This gives us $\vec z(\vec R,g_s)$ that we then use in order to get $\vec y(\vec R_0,g_s,\vec z(\vec R,g_s))$. Since we assume here that there is no new physics, our predictions
\begin{flalign}
\lim_{L\mapsto \infty,g_s\mapsto 0} \sigma\left(g_s,\vec y(\vec R_0,g_s,\vec z(\vec R,g_s)),\vec z(\vec R,g_s),L\right) = \sigma^\mathrm{Nature}
\label{eq:nature}
\end{flalign}
will not depend upon the choice of the experimental quantities $\vec R$ that we used in the matching procedure.

At this point we introduce an additional step, the \emph{definition} of isoQCD. Since we want to be fully general, we consider a new iso-symmetric input vector, let's call it $\vec S_0$. The entries of this vector may or may not coincide with $\vec R_0$, may be the experimental values of measurable quantities or theory scales. By using this input vector we implement the first step of the matching algorithm with $\vec z=\vec 0$,
\begin{flalign}
&
\lim_{L\mapsto \infty}\vec S_0\left(g_s,\vec y,\vec 0,L\right) = \vec S_0\;,
\quad \longmapsto \quad 
\vec y(\vec S_0,g_s,\vec 0)\;.
\label{eq:twosteps1iso}
\end{flalign}

Once we know the iso-symmetric quark masses and the iso-symmetric lattice spacing we can compute any observable in isoQCD, e.g.
\begin{flalign}
\lim_{L\mapsto \infty,g_s\mapsto 0} \sigma\left(g_s,\vec y(\vec S_0,g_s,\vec 0),\vec 0,L\right) = \sigma^\mathrm{isoQCD}(\vec S_0)\;.
\label{eq:continuumiso}
\end{flalign}
Our isoQCD prediction $\sigma^\mathrm{isoQCD}(\vec S_0)$  \emph{does} depend upon our choice of the input vector $\vec S_0$ and, consequently, also the QED and SIB corrections
\begin{flalign}
\Delta \sigma^\mathrm{IB}(\vec S_0)\equiv \sigma^\mathrm{Nature} - \sigma^\mathrm{isoQCD}(\vec S_0)
\end{flalign}
\emph{do} depend upon the prescription that we adopted to define isoQCD.  

Although there is no theoretical constraint on the input vector $\vec S_0$, a totally arbitrary choice cannot correspond to a definition of isoQCD that is a good approximation of the hadronic Universe. In practice we have to use quantities that we know from experiments or from a theoretical QCD$+$QED calculation that we, or a friend of us, did in the past. For example, we could use the same quantities defining the vector $\vec R_0$ in Eq.~(\ref{eq:RR0R1}) but, since in isoQCD all pions and kaons have the same mass and are stable, we could e.g.\ use $M_{\pi^0}$ instead of $M_{\pi^+}$ and $(M_{K^0}+M_{K^+})/2$ instead of $M_{K^0}$. A moment of thought reveals that, in fact, there is no loss of generality if we write
\begin{flalign}
\vec S_0\equiv \vec R_0+\vec \varepsilon\;,
\qquad
\vec \varepsilon=\mathcal{O}(\vec z)\;,
\end{flalign}
where $\vec \varepsilon$ is a correction vector with entries of the order of the experimentally measured IB corrections to the hadron masses defining $\vec R_0$. Indeed, as discussed in subsection~\ref{sec:afmatching}, the same line of constant physics corresponding to any given choice of $\vec S_0$ can also be selected by using $\vec R_0^\mathrm{isoQCD}(\vec S_0)\equiv \vec R_0+\vec \varepsilon$.

\subsection{\`a la RM123}
As stressed several times, \emph{leading} QED and SIB corrections must be taken into account at the sub-percent level of accuracy. At the same time, sub-leading corrections can safely be neglected. Moreover, leading IB corrections can be directly computed by using the so-called RM123 approach~\cite{deDivitiis:2011eh,deDivitiis:2013xla}, i.e.\ by expanding the lattice path integral w.r.t.\ the IB couplings $\vec z$. For these reasons it is useful and instructive to elaborate a bit more on the definition of isoQCD, and of the associated QED and SIB corrections, by expanding the previous formulae w.r.t.\ $\vec z$ and by neglecting terms of $\mathcal{O}(\vec z^2)$. This is what we are now going to do.

Let's start once again from QCD$+$QED, more precisely from Eq.~(\ref{eq:nature}) that we linearize w.r.t.\ $\vec z$,
\begin{flalign}
\lim_{g_s\mapsto 0} 
\left\{
\sigma\left(g_s,\vec y(\vec R_0,g_s,\vec z(\vec R,g_s)),\vec 0\right)
+
z^i(\vec R,g_s) 
\left.
\frac{\partial \sigma\left(g_s,\vec y(\vec R_0,g_s,\vec z(\vec R,g_s)),\vec z\right)}{\partial z^i}
\right\vert_{\vec z=\vec 0}
\right\}
= \sigma^\mathrm{Nature}\;.
\label{eq:linnature}
\end{flalign}
In the previous expression, and also in the ones that will follow, we neglect $\mathcal{O}(\vec z^2)$ contributions and, to simplify the notation, we assume that the infinite-volume limit has already been taken. 

A very important remark is in order here. It is not possible to take, separately, the continuum limit of the two terms in curly brackets on the l.h.s.\ of Eq.~(\ref{eq:linnature}). Indeed,
\begin{flalign}
\lim_{g_s\mapsto 0} 
\sigma\left(g_s,\vec y(\vec R_0,g_s,\vec z(\vec R,g_s)),\vec 0\right)
=\infty\;!
\end{flalign}
The different UV behaviors of isoQCD and QCD$+$QED come into play here. Both theories are renormalizable and the different UV-divergences are absorbed into the different bare parameters, resulting in turn from the different matching conditions. In order to perform, separately, the continuum limits of the isoQCD contribution to our observable $\sigma$ and of the associated QED and SIB corrections we must use the isoQCD bare parameters and the associated \emph{counter-terms},
\begin{flalign}
\Delta \vec y 
= \vec y\left(\vec R_0,g_s,\vec z(\vec R,g_s)\right)-\vec y(\vec R_0+\varepsilon,g_s,\vec 0)\;,
\label{eq:counterterms}
\end{flalign}
i.e.\ the difference between the IB bare parameters of full theory and those of isoQCD. By working \`a la RM123, the counter-terms can be obtained by linearizing the matching conditions of Eq.~(\ref{eq:twosteps1}) and Eq.~(\ref{eq:twosteps2}). 
Once this has been done we can go back to Eq.~(\ref{eq:linnature}) and split it as follows
\begin{flalign}
&
\sigma^\mathrm{isoQCD}(\vec R_0+\varepsilon)
=\lim_{g_s\mapsto 0} 
\sigma\left(g_s,\vec y(\vec R_0+\varepsilon,g_s,\vec 0),\vec 0\right)\;,
\end{flalign}
and
\begin{flalign}
&
\Delta \sigma^\mathrm{IB}(\vec R_0+\varepsilon)
\nonumber \\
&=\lim_{g_s\mapsto 0} 
\left\{
\Delta y^n 
\left.
\frac{\partial \sigma\left(g_s,\vec y,\vec 0\right)}{\partial y^n}
\right\vert_{\vec y=\vec y(\vec R_0+\varepsilon,g_s,\vec 0)}
+
z^i(\vec R,g_s) 
\left.
\frac{\partial \sigma\left(g_s,\vec y(\vec R_0+\varepsilon,g_s,\vec 0),\vec z\right)}{\partial z^i}
\right\vert_{\vec z=\vec 0}
\right\}\;,
\label{eq:linnaturesplit}
\end{flalign}
where both $\sigma^\mathrm{isoQCD}(\vec R_0+\varepsilon)$ and $\Delta \sigma^\mathrm{IB}(\vec R_0+\varepsilon)$ have a well defined continuum limit (see Refs.\cite{deDivitiis:2013xla,Tantalo:2013maa}). 

At this point I really have to apologize with the reader that has been coping with a very very heavy notation! The formulae above could have been written by partly hiding the dependence upon the inputs and/or by giving-up the full generality that we insisted in retaining. At the end of this climb, however, we have to be pretty satisfied. Indeed, by using our formulae we can now express the dependence of isoQCD observables w.r.t.\ the prescription used to define the approximate theory in terms of the differential equations
\begin{flalign}
\sigma^\mathrm{isoQCD}(\vec R_0+\varepsilon) -\sigma^\mathrm{isoQCD}(\vec R_0)
=\varepsilon^n \frac{\partial \sigma^\mathrm{isoQCD}(\vec R_0)}{\partial R_0^n} \;,
\label{eq:prescrpdep1}
\end{flalign}
i.e.\ in terms of the partial derivatives of our observable $\sigma$ w.r.t.\ the physical inputs $\vec R_0$.

At the same time, by observing that we can rewrite Eq.~(\ref{eq:counterterms}) as
\begin{flalign}
\Delta \vec y 
= 
-\varepsilon_n \frac{\partial \vec y(\vec R_0,g_s,\vec 0)}{\partial R_0^n}
+z^i(\vec R,g_s) 
\left.\frac{\partial \vec y(\vec R_0,g_s,\vec z)}{\partial z^i}\right\vert_{\vec z=\vec 0}
=\mathcal{O}(\vec z)
\;,
\label{eq:counterterms2}
\end{flalign}
we can safely replace $\vec y(\vec R_0+\varepsilon,g_s,\vec 0)$ with $\vec y(\vec R_0,g_s,\vec 0)$ in the expression for $\Delta \sigma^\mathrm{IB}(\vec R_0+\varepsilon)$ given in Eq.~(\ref{eq:linnaturesplit}) and rewrite it as follows
\begin{flalign}
&
\Delta \sigma^\mathrm{IB}(\vec R_0+\varepsilon)
=
- \varepsilon^n \frac{\partial \sigma^\mathrm{isoQCD}(\vec R_0)}{\partial R_0^n}
+
\Delta \sigma^\mathrm{IB}(\vec R_0)
\;.
\label{eq:linnaturesplit2}
\end{flalign}
This expression, that we obtained by using
\begin{flalign}
\lim_{g_s\mapsto 0} 
\frac{\partial y^m(\vec R_0,g_s,\vec 0)}{\partial R_0^n}
\frac{\partial \sigma(g_s,\vec y(\vec R_0,g_s,\vec 0),\vec 0)}{\partial y^m(\vec R_0,g_s,\vec 0)}
= \frac{\partial \sigma^\mathrm{isoQCD}(\vec R_0)}{\partial R_0^n} \;,
\end{flalign}
shows explicitly what we was expecting: the dependence upon the prescription of an isoQCD observable (see Eq.~(\ref{eq:prescrpdep1})) is exactly compensated by the dependence of the corresponding IB correction (see Eq.~(\ref{eq:linnaturesplit2})), so that their sum is the physical quantity $\sigma^\mathrm{Nature}$
independently from the choices made in order to define isoQCD!

Moreover, by noticing that
\begin{flalign}
&
\Delta \sigma^\mathrm{IB}(\vec R_0)
=
\lim_{g_s\mapsto 0} 
z^i(\vec R,g_s)\,
\left.
\frac{\partial \sigma\left(g_s,\vec y(\vec R_0,g_s,\vec z),\vec z\right)}{\partial z^i}
\right\vert_{\vec z=\vec 0}
\;,
\label{eq:linnaturesplit3}
\end{flalign}
we see explicitly that, if the same external (and experimental) inputs are used in order to match isoQCD and QCD$+$QED, at fixed UV cutoff the resulting IB correction is nothing but the leading term in the Taylor expansion of our observable w.r.t.\ the IB bare parameters $e^2$ and $(am_u-am_d)/2$ (see Eq.~(\ref{eq:splittingbare})).

\section{Does the choice of the QCD prescription matter?}
In principle it does. In the previous section we learned that
\begin{flalign}
\sigma^\mathrm{isoQCD}(\vec R_0+\varepsilon) -\sigma^\mathrm{isoQCD}(\vec R_0)
=\varepsilon^n \frac{\partial \sigma^\mathrm{isoQCD}(\vec R_0)}{\partial R_0^n} =\mathcal{O}(\vec \varepsilon)\;,
\end{flalign}
i.e.\ the difference between isoQCD results obtained with different prescriptions is of the same order of the IB corrections. If the IB corrections do matter also the scheme dependence is important. 

In practice, as rarely happens in life, we are quite lucky. Although different lattice collaborations adopted different prescriptions to define isoQCD no significant differences have been observed yet within the quoted uncertainties. 

\begin{figure}[t!]
\begin{center}
\includegraphics[width=0.8\textwidth]{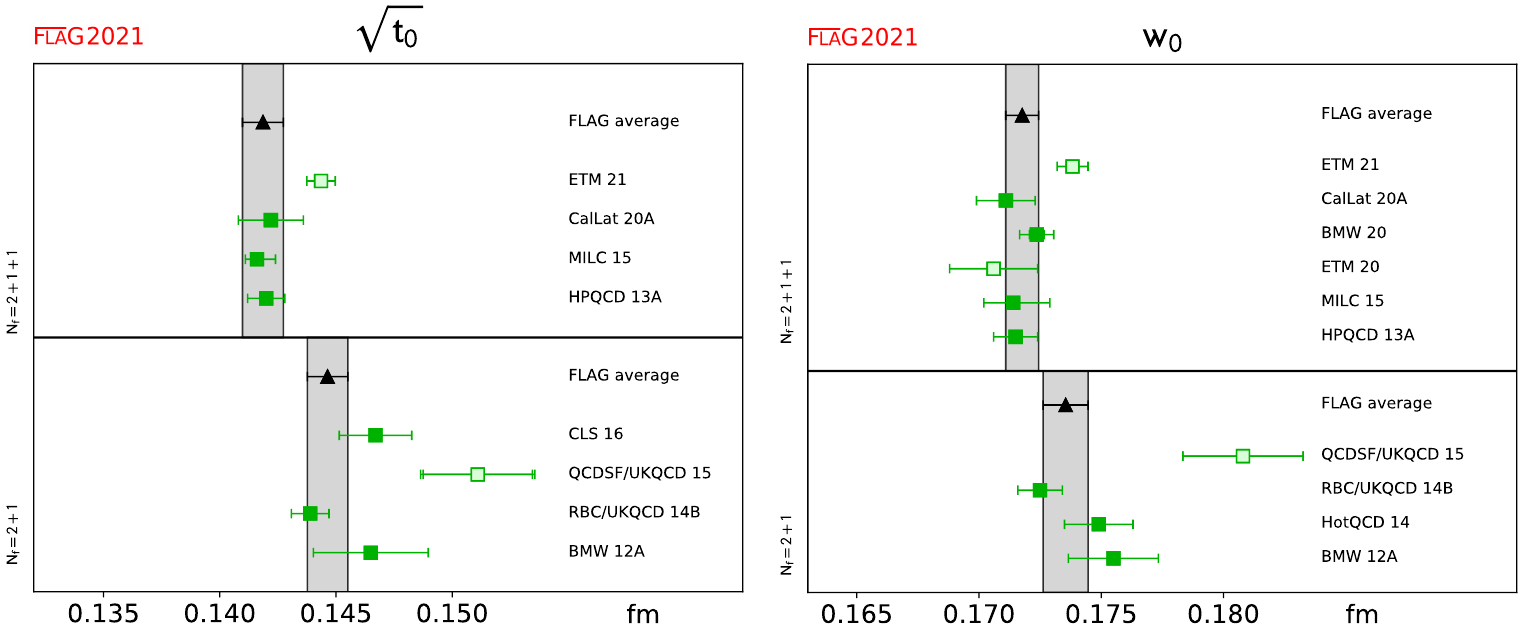}
\end{center}
\caption{\label{fig:FLAG} FLAG~\cite{FlavourLatticeAveragingGroupFLAG:2021npn} average of the lattice determinations of the gradient-flow scales $t_0$ and $w_0$.}
\end{figure}

We cannot expect this to be true for any observable but, for example, we can consider the gradient-flow theory scales $t_0$~\cite{Luscher:2010iy} and $w_0$~\cite{Borsanyi:2012zs} that can be computed quite precisely on the lattice. This has been done in Ref.~\cite{FlavourLatticeAveragingGroupFLAG:2021npn} by the FLAG scale-setting working group, see Figure~\ref{fig:FLAG}. 
The different collaborations~\cite{ExtendedTwistedMass:2021qui,Miller:2020evg,Borsanyi:2020mff,ExtendedTwistedMass:2020tvp,MILC:2015tqx,Dowdall:2013rya,Bruno:2016plf,Bornyakov:2015eaa,RBC:2014ntl,HotQCD:2014kol,Borsanyi:2012zs} used different observables to define isoQCD (see Table~76 of Ref.~\cite{FlavourLatticeAveragingGroupFLAG:2021npn}) and also different values for these observables. In most of the cases $M_{\pi^0}^\mathrm{exp}$ has been used but different values, in the range $[494.2,497.6]$~MeV, have been adopted for the kaon mass. Leptonic decay constants have been used in Refs.~\cite{ExtendedTwistedMass:2021qui,ExtendedTwistedMass:2020tvp,MILC:2015tqx,Dowdall:2013rya,Bruno:2016plf,HotQCD:2014kol} while $M_{\Omega^-}^\mathrm{exp}$ has been adopted in Refs.~\cite{Miller:2020evg,Borsanyi:2020mff,RBC:2014ntl,Borsanyi:2012zs}. In fact, a careful analysis reveals that the differences observed in Figure~\ref{fig:FLAG} cannot be ascribed to the scheme dependence.

\begin{figure}[t!]
\begin{center}
\includegraphics[width=0.8\textwidth]{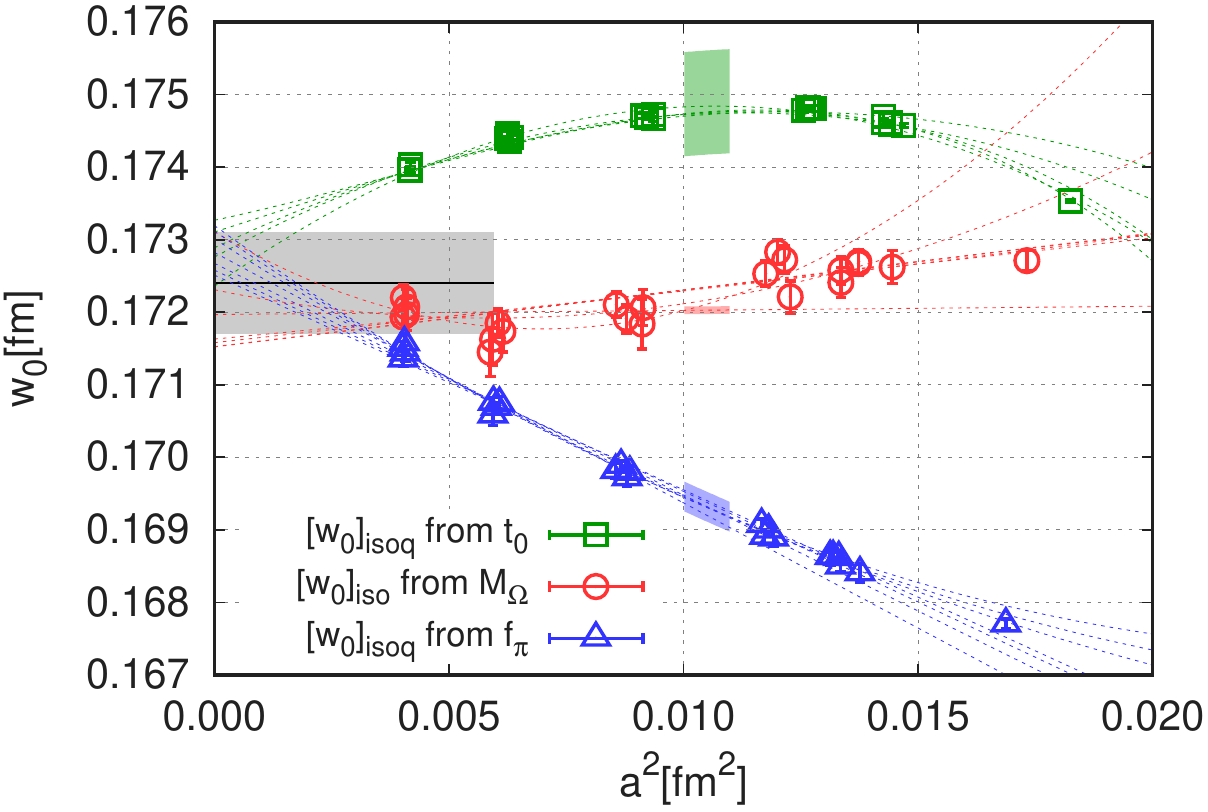}
\end{center}
\caption{\label{fig:BMW} Study of the dependence of $w_0$ upon the prescription used to define isoQCD performed in Ref.~\cite{Borsanyi:2020mff}}
\end{figure}
In the case of $w_0$ a dedicated study of the scheme dependence has been performed in Ref.~\cite{Borsanyi:2020mff}, from which we extracted Figure~\ref{fig:BMW}. The figure shows the continuum extrapolation of $w_0$ obtained by using different inputs to define isoQCD. More precisely, the red data have been obtained by using $M_{\pi^0}^\mathrm{exp}$, the theory scale $M^{\Gamma_5}_{ss}=689.89(49)$~MeV and $M_{\Omega^-}^\mathrm{exp}$. The blue and green data have been obtained by using $M_{\pi}=134.8(3)$~MeV, $M_{K}=494.2(3)$~MeV, $f_\pi=130.50(14)$~MeV (blue) and $t_0=0.1416(^{+8}_{-5})$~fm (green). Although the data corresponding to different prescriptions do differ at fixed UV cutoff, no significant differences are observed within the quoted errors in the continuum.

A detailed numerical investigation of the scheme dependence has also been performed in Refs.~\cite{DiCarlo:2019thl,Boyle:2022lsi} in the case of the leptonic decay rates of pions and kaons. In Ref.~\cite{DiCarlo:2019thl} it has been demonstrated the numerical equivalence, at the current level of precision, of the hadronic scheme defined by $M_\pi=135.0(2)$~MeV, $M_K=494.6(1)$~MeV, $M_{D_s}=1966.7(1.5)$~MeV, $f_\pi=130.65(12)$~MeV with the so-called GRS scheme~\cite{Gasser:2003hk}. The GRS scheme is particularly attractive because defines isoQCD by imposing that the renormalized couplings (theory scales)
\begin{flalign}
&
\hat{g}_s(\mu),\ \frac{\hat m_u(\mu)+\hat m_d(\mu)}{2},\ \hat m_s(\mu),\ \hat m_c(\mu)
\end{flalign}
are the same in isoQCD and in QCD$+$QED and, therefore, matches the two theories by using short-distance conditions that can easily be implemented in analytical calculations (e.g.\ in high-energy perturbative calculations or in chiral perturbation theory). In fact one has different GRS schemes depending upon the choices of the renormalization scheme and of the scale $\mu$ and, as originally done in Ref.~\cite{deDivitiis:2013xla}, $\mu=2$~GeV and the $\overline{\mathrm{MS}}$ scheme have been used in Ref.~\cite{DiCarlo:2019thl}. A similar study, albeit at a single value of the lattice spacing, has been performed more recently in Ref.~\cite{Boyle:2022lsi}.

As I said at the beginning of this section, we have been quite lucky so far. Although we must expect differences between isoQCD results obtained in different schemes, in fact, the adopted schemes are not as different as it might appear at first sight!

\section{The past is the past, what about the future?}
Thank God a consensus is emerging within the lattice community on the necessity of agreeing on a scheme to define isoQCD. That's why I have been invited to give this talk in which I have been mostly discussing well established facts about renormalization theory. 

I'm not going to make an explicit proposal here for the scheme to be used. We could use physical quantities and their experimental measurements, slightly corrected experimental values, theory scales, whatever. What really matters is that we sit down around a table and decide what has to be done in the future. An explicit proposal will be presented in the next edition of the FLAG review. Stay tuned and, please, take it in due consideration! 

\acknowledgments{I thank the organizers of the conference for honoring me with the opportunity of discussing this important topic. I warmly thank my colleagues of the RM123, RC$^\star$ and FLAG collaborations. The material for this talk stems from the countless, endless, and delightful discussions I've had with these friends of mine over the past decade.}



\begin{thebibliography}{99}
\bibitem{FlavourLatticeAveragingGroupFLAG:2021npn}
Y.~Aoki \textit{et al.} [Flavour Lattice Averaging Group (FLAG)],
Eur. Phys. J. C \textbf{82} (2022) no.10, 869
doi:10.1140/epjc/s10052-022-10536-1
[arXiv:2111.09849 [hep-lat]].

\bibitem{Duncan:1996xy}
A.~Duncan, E.~Eichten and H.~Thacker,
Phys. Rev. Lett. \textbf{76} (1996), 3894-3897
doi:10.1103/PhysRevLett.76.3894
[arXiv:hep-lat/9602005 [hep-lat]].

\bibitem{Blum:2007cy}
T.~Blum, T.~Doi, M.~Hayakawa, T.~Izubuchi and N.~Yamada,
Phys. Rev. D \textbf{76} (2007), 114508
doi:10.1103/PhysRevD.76.114508
[arXiv:0708.0484 [hep-lat]].

\bibitem{Hayakawa:2008an}
M.~Hayakawa and S.~Uno,
Prog. Theor. Phys. \textbf{120} (2008), 413-441
doi:10.1143/PTP.120.413
[arXiv:0804.2044 [hep-ph]].

\bibitem{Blum:2010ym}
T.~Blum, R.~Zhou, T.~Doi, M.~Hayakawa, T.~Izubuchi, S.~Uno and N.~Yamada,
Phys. Rev. D \textbf{82} (2010), 094508
doi:10.1103/PhysRevD.82.094508
[arXiv:1006.1311 [hep-lat]].

\bibitem{deDivitiis:2011eh}
G.~M.~de Divitiis, P.~Dimopoulos, R.~Frezzotti, V.~Lubicz, G.~Martinelli, R.~Petronzio, G.~C.~Rossi, F.~Sanfilippo, S.~Simula and N.~Tantalo, \textit{et al.}
JHEP \textbf{04} (2012), 124
doi:10.1007/JHEP04(2012)124
[arXiv:1110.6294 [hep-lat]].

\bibitem{deDivitiis:2013xla}
G.~M.~de Divitiis \textit{et al.} [RM123],
Phys. Rev. D \textbf{87} (2013) no.11, 114505
doi:10.1103/PhysRevD.87.114505
[arXiv:1303.4896 [hep-lat]].

\bibitem{Budapest-Marseille-Wuppertal:2013rtp}
S.~Borsanyi \textit{et al.} [Budapest-Marseille-Wuppertal],
Phys. Rev. Lett. \textbf{111} (2013) no.25, 252001
doi:10.1103/PhysRevLett.111.252001
[arXiv:1306.2287 [hep-lat]].

\bibitem{Davoudi:2014qua}
Z.~Davoudi and M.~J.~Savage,
Phys. Rev. D \textbf{90} (2014) no.5, 054503
doi:10.1103/PhysRevD.90.054503
[arXiv:1402.6741 [hep-lat]].

\bibitem{Borsanyi:2014jba}
S.~Borsanyi, S.~Durr, Z.~Fodor, C.~Hoelbling, S.~D.~Katz, S.~Krieg, L.~Lellouch, T.~Lippert, A.~Portelli and K.~K.~Szabo, \textit{et al.}
Science \textbf{347} (2015), 1452-1455
doi:10.1126/science.1257050
[arXiv:1406.4088 [hep-lat]].

\bibitem{Lucini:2015hfa}
B.~Lucini, A.~Patella, A.~Ramos and N.~Tantalo,
JHEP \textbf{02} (2016), 076
doi:10.1007/JHEP02(2016)076
[arXiv:1509.01636 [hep-th]].

\bibitem{Horsley:2015vla}
R.~Horsley, Y.~Nakamura, H.~Perlt, D.~Pleiter, P.~E.~L.~Rakow, G.~Schierholz, A.~Schiller, R.~Stokes, H.~St\"uben and R.~D.~Young, \textit{et al.}
JHEP \textbf{04} (2016), 093
doi:10.1007/JHEP04(2016)093
[arXiv:1509.00799 [hep-lat]].

\bibitem{Horsley:2015eaa}
R.~Horsley, Y.~Nakamura, H.~Perlt, D.~Pleiter, P.~E.~L.~Rakow, G.~Schierholz, A.~Schiller, R.~Stokes, H.~St\"uben and R.~D.~Young, \textit{et al.}
J. Phys. G \textbf{43} (2016) no.10, 10LT02
doi:10.1088/0954-3899/43/10/10LT02
[arXiv:1508.06401 [hep-lat]].

\bibitem{Endres:2015gda}
M.~G.~Endres, A.~Shindler, B.~C.~Tiburzi and A.~Walker-Loud,
Phys. Rev. Lett. \textbf{117} (2016) no.7, 072002
doi:10.1103/PhysRevLett.117.072002
[arXiv:1507.08916 [hep-lat]].

\bibitem{Carrasco:2015xwa}
N.~Carrasco, V.~Lubicz, G.~Martinelli, C.~T.~Sachrajda, N.~Tantalo, C.~Tarantino and M.~Testa,
Phys. Rev. D \textbf{91} (2015) no.7, 074506
doi:10.1103/PhysRevD.91.074506
[arXiv:1502.00257 [hep-lat]].

\bibitem{Lee:2015rua}
J.~W.~Lee and B.~C.~Tiburzi,
Phys. Rev. D \textbf{93} (2016) no.3, 034012
doi:10.1103/PhysRevD.93.034012
[arXiv:1508.04165 [hep-lat]].

\bibitem{Tantalo:2016vxk}
N.~Tantalo, V.~Lubicz, G.~Martinelli, C.~T.~Sachrajda, F.~Sanfilippo and S.~Simula,
[arXiv:1612.00199 [hep-lat]].

\bibitem{Lubicz:2016xro}
V.~Lubicz, G.~Martinelli, C.~T.~Sachrajda, F.~Sanfilippo, S.~Simula and N.~Tantalo,
Phys. Rev. D \textbf{95} (2017) no.3, 034504
doi:10.1103/PhysRevD.95.034504
[arXiv:1611.08497 [hep-lat]].

\bibitem{Blum:2017cer}
T.~Blum, N.~Christ, M.~Hayakawa, T.~Izubuchi, L.~Jin, C.~Jung and C.~Lehner,
Phys. Rev. D \textbf{96} (2017) no.3, 034515
doi:10.1103/PhysRevD.96.034515
[arXiv:1705.01067 [hep-lat]].

\bibitem{Giusti:2017dmp}
D.~Giusti, V.~Lubicz, C.~Tarantino, G.~Martinelli, F.~Sanfilippo, S.~Simula and N.~Tantalo,
Phys. Rev. D \textbf{95} (2017) no.11, 114504
doi:10.1103/PhysRevD.95.114504
[arXiv:1704.06561 [hep-lat]].

\bibitem{Patella:2017fgk}
A.~Patella,
PoS \textbf{LATTICE2016} (2017), 020
doi:10.22323/1.256.0020
[arXiv:1702.03857 [hep-lat]].

\bibitem{Boyle:2017gzv}
P.~Boyle, V.~G\"ulpers, J.~Harrison, A.~J\"uttner, C.~Lehner, A.~Portelli and C.~T.~Sachrajda,
JHEP \textbf{09} (2017), 153
doi:10.1007/JHEP09(2017)153
[arXiv:1706.05293 [hep-lat]].

\bibitem{Bussone:2017xkb}
A.~Bussone, M.~Della Morte and T.~Janowski,
EPJ Web Conf. \textbf{175} (2018), 06005
doi:10.1051/epjconf/201817506005
[arXiv:1710.06024 [hep-lat]].

\bibitem{Feng:2018qpx}
X.~Feng and L.~Jin,
Phys. Rev. D \textbf{100} (2019) no.9, 094509
doi:10.1103/PhysRevD.100.094509
[arXiv:1812.09817 [hep-lat]].

\bibitem{Davoudi:2018qpl}
Z.~Davoudi, J.~Harrison, A.~J\"uttner, A.~Portelli and M.~J.~Savage,
Phys. Rev. D \textbf{99} (2019) no.3, 034510
doi:10.1103/PhysRevD.99.034510
[arXiv:1810.05923 [hep-lat]].

\bibitem{MILC:2018ddw}
S.~Basak \textit{et al.} [MILC],
Phys. Rev. D \textbf{99} (2019) no.3, 034503
doi:10.1103/PhysRevD.99.034503
[arXiv:1807.05556 [hep-lat]].

\bibitem{Bijnens:2019ejw}
J.~Bijnens, J.~Harrison, N.~Hermansson-Truedsson, T.~Janowski, A.~J\"uttner and A.~Portelli,
Phys. Rev. D \textbf{100} (2019) no.1, 014508
doi:10.1103/PhysRevD.100.014508
[arXiv:1903.10591 [hep-lat]].

\bibitem{CSSM:2019jmq}
R.~Horsley \textit{et al.} [CSSM, QCDSF and UKQCD],
J. Phys. G \textbf{46} (2019), 115004
doi:10.1088/1361-6471/ab32c1
[arXiv:1904.02304 [hep-lat]].

\bibitem{Giusti:2019xct}
D.~Giusti, V.~Lubicz, G.~Martinelli, F.~Sanfilippo and S.~Simula,
Phys. Rev. D \textbf{99} (2019) no.11, 114502
doi:10.1103/PhysRevD.99.114502
[arXiv:1901.10462 [hep-lat]].

\bibitem{Hatton:2020qhk}
D.~Hatton \textit{et al.} [HPQCD],
Phys. Rev. D \textbf{102} (2020) no.5, 054511
doi:10.1103/PhysRevD.102.054511
[arXiv:2005.01845 [hep-lat]].

\bibitem{DiCarlo:2021apt}
M.~Di Carlo, M.~T.~Hansen, A.~Portelli and N.~Hermansson-Truedsson,
Phys. Rev. D \textbf{105} (2022) no.7, 074509
doi:10.1103/PhysRevD.105.074509
[arXiv:2109.05002 [hep-lat]].

\bibitem{Feng:2021zek}
X.~Feng, L.~Jin and M.~J.~Riberdy,
Phys. Rev. Lett. \textbf{128} (2022) no.5, 052003
doi:10.1103/PhysRevLett.128.052003
[arXiv:2108.05311 [hep-lat]].

\bibitem{Hatton:2020lnm}
D.~Hatton, C.~T.~H.~Davies and G.~P.~Lepage,
Phys. Rev. D \textbf{102} (2020) no.9, 094514
doi:10.1103/PhysRevD.102.094514
[arXiv:2009.07667 [hep-lat]].

\bibitem{Aoki:2012st}
S.~Aoki, K.~I.~Ishikawa, N.~Ishizuka, K.~Kanaya, Y.~Kuramashi, Y.~Nakamura, Y.~Namekawa, M.~Okawa, Y.~Taniguchi and A.~Ukawa, \textit{et al.}
Phys. Rev. D \textbf{86} (2012), 034507
doi:10.1103/PhysRevD.86.034507
[arXiv:1205.2961 [hep-lat]].

\bibitem{FermilabLattice:2021hzx}
C.~McNeile \textit{et al.} [Fermilab Lattice, HPQCD and MILC],
PoS \textbf{LATTICE2021} (2022), 039
doi:10.22323/1.396.0039
[arXiv:2112.11339 [hep-lat]].

\bibitem{Gasser:2003hk}
J.~Gasser, A.~Rusetsky and I.~Scimemi,
Eur. Phys. J. C \textbf{32} (2003), 97-114
doi:10.1140/epjc/s2003-01383-1
[arXiv:hep-ph/0305260 [hep-ph]].

\bibitem{Bussone:2018ybs}
A.~Bussone, M.~Della Morte, T.~Janowski and A.~Walker-Loud,
PoS \textbf{LATTICE2018} (2018), 293
doi:10.22323/1.334.0293
[arXiv:1810.11647 [hep-lat]].

\bibitem{DiCarlo:2019thl}
M.~Di Carlo, D.~Giusti, V.~Lubicz, G.~Martinelli, C.~T.~Sachrajda, F.~Sanfilippo, S.~Simula and N.~Tantalo,
Phys. Rev. D \textbf{100} (2019) no.3, 034514
doi:10.1103/PhysRevD.100.034514
[arXiv:1904.08731 [hep-lat]].

\bibitem{Borsanyi:2020mff}
S.~Borsanyi, Z.~Fodor, J.~N.~Guenther, C.~Hoelbling, S.~D.~Katz, L.~Lellouch, T.~Lippert, K.~Miura, L.~Parato and K.~K.~Szabo, \textit{et al.}
Nature \textbf{593} (2021) no.7857, 51-55
doi:10.1038/s41586-021-03418-1
[arXiv:2002.12347 [hep-lat]].

\bibitem{Sachrajda:2021enz}
C.~T.~Sachrajda,
Acta Phys. Polon. B \textbf{52} (2021) no.3, 175-201
doi:10.5506/APhysPolB.52.175
[arXiv:2104.04312 [hep-lat]].

\bibitem{Boyle:2022lsi}
P.~Boyle, M.~Di Carlo, F.~Erben, V.~G\"ulpers, M.~T.~Hansen, T.~Harris, N.~Hermansson-Truedsson, R.~Hodgson, A.~J\"uttner and F.~\'O.~h\'Og\'ain, \textit{et al.}
[arXiv:2211.12865 [hep-lat]].

\bibitem{Luscher:2010iy}
M.~L\"uscher,
JHEP \textbf{08} (2010), 071
[erratum: JHEP \textbf{03} (2014), 092]
doi:10.1007/JHEP08(2010)071
[arXiv:1006.4518 [hep-lat]].

\bibitem{Borsanyi:2012zs}
S.~Borsanyi, S.~Durr, Z.~Fodor, C.~Hoelbling, S.~D.~Katz, S.~Krieg, T.~Kurth, L.~Lellouch, T.~Lippert and C.~McNeile, \textit{et al.}
JHEP \textbf{09} (2012), 010
doi:10.1007/JHEP09(2012)010
[arXiv:1203.4469 [hep-lat]].

\bibitem{Bloch:1937pw}
F.~Bloch and A.~Nordsieck,
Phys. Rev. \textbf{52} (1937), 54-59
doi:10.1103/PhysRev.52.54

\bibitem{Giusti:2017dwk}
D.~Giusti, V.~Lubicz, G.~Martinelli, C.~T.~Sachrajda, F.~Sanfilippo, S.~Simula, N.~Tantalo and C.~Tarantino,
Phys. Rev. Lett. \textbf{120} (2018) no.7, 072001
doi:10.1103/PhysRevLett.120.072001
[arXiv:1711.06537 [hep-lat]].

\bibitem{Hansen:2018zre}
M.~Hansen, B.~Lucini, A.~Patella and N.~Tantalo,
JHEP \textbf{05} (2018), 146
doi:10.1007/JHEP05(2018)146
[arXiv:1802.05474 [hep-lat]].

\bibitem{Bushnaq:2022aam}
L.~Bushnaq, I.~Campos, M.~Catillo, A.~Cotellucci, M.~Dale, P.~Fritzsch, J.~L\"ucke, M.~Krsti\'c Marinkovi\'c, A.~Patella and N.~Tantalo,
[arXiv:2209.13183 [hep-lat]].

\bibitem{DelDebbio:2022uve}
L.~Del Debbio, A.~Lupo, M.~Panero and N.~Tantalo,
[arXiv:2212.08019 [hep-lat]].

\bibitem{Hansen:2019idp}
M.~Hansen, A.~Lupo and N.~Tantalo,
Phys. Rev. D \textbf{99} (2019) no.9, 094508
doi:10.1103/PhysRevD.99.094508
[arXiv:1903.06476 [hep-lat]].

\bibitem{RBC:2014ntl}
T.~Blum \textit{et al.} [RBC and UKQCD],
Phys. Rev. D \textbf{93} (2016) no.7, 074505
doi:10.1103/PhysRevD.93.074505
[arXiv:1411.7017 [hep-lat]].

\bibitem{Segner:2022dou}
A.~M.~Segner, A.~D.~Hanlon, A.~Risch and H.~Wittig,
[arXiv:2212.07176 [hep-lat]].

\bibitem{RQCD:2022xux}
G.~S.~Bali \textit{et al.} [RQCD],
[arXiv:2211.03744 [hep-lat]].

\bibitem{Tantalo:2013maa}
N.~Tantalo,
PoS \textbf{LATTICE2013} (2014), 007
doi:10.22323/1.187.0007
[arXiv:1311.2797 [hep-lat]].

\bibitem{ExtendedTwistedMass:2021qui}
C.~Alexandrou \textit{et al.} [Extended Twisted Mass],
Phys. Rev. D \textbf{104} (2021) no.7, 074520
doi:10.1103/PhysRevD.104.074520
[arXiv:2104.06747 [hep-lat]].

\bibitem{Miller:2020evg}
N.~Miller, L.~Carpenter, E.~Berkowitz, C.~C.~Chang, B.~H\"orz, D.~Howarth, H.~Monge-Camacho, E.~Rinaldi, D.~A.~Brantley and C.~K\"orber, \textit{et al.}
Phys. Rev. D \textbf{103} (2021) no.5, 054511
doi:10.1103/PhysRevD.103.054511
[arXiv:2011.12166 [hep-lat]].

\bibitem{ExtendedTwistedMass:2020tvp}
G.~Bergner \textit{et al.} [Extended Twisted Mass],
PoS \textbf{LATTICE2019} (2020), 181
doi:10.22323/1.363.0181
[arXiv:2001.09116 [hep-lat]].

\bibitem{MILC:2015tqx}
A.~Bazavov \textit{et al.} [MILC],
Phys. Rev. D \textbf{93} (2016) no.9, 094510
doi:10.1103/PhysRevD.93.094510
[arXiv:1503.02769 [hep-lat]].

\bibitem{Dowdall:2013rya}
R.~J.~Dowdall, C.~T.~H.~Davies, G.~P.~Lepage and C.~McNeile,
Phys. Rev. D \textbf{88} (2013), 074504
doi:10.1103/PhysRevD.88.074504
[arXiv:1303.1670 [hep-lat]].

\bibitem{Bruno:2016plf}
M.~Bruno, T.~Korzec and S.~Schaefer,
Phys. Rev. D \textbf{95} (2017) no.7, 074504
doi:10.1103/PhysRevD.95.074504
[arXiv:1608.08900 [hep-lat]].

\bibitem{Bornyakov:2015eaa}
V.~G.~Bornyakov, R.~Horsley, R.~Hudspith, Y.~Nakamura, H.~Perlt, D.~Pleiter, P.~E.~L.~Rakow, G.~Schierholz, A.~Schiller and H.~St\"uben, \textit{et al.}
[arXiv:1508.05916 [hep-lat]].

\bibitem{HotQCD:2014kol}
A.~Bazavov \textit{et al.} [HotQCD],
Phys. Rev. D \textbf{90} (2014), 094503
doi:10.1103/PhysRevD.90.094503
[arXiv:1407.6387 [hep-lat]].
\end{thebibliography}
\end{document}